\documentclass[sigconf,10pt,nonacm]{acmart}

\title{Online Advertising is a Regrettable Necessity: On the Dangers of Pay-Walling the Web}
\author{
    Yonas Kassa \\
}
\email{ykassa@acm.org}

\begin{abstract}
The exponential growth of the web and its benefits can be attributed largely to its open model where anyone with internet connection can access information on the web for free. This has created unprecedented opportunities for various members of society including the most vulnerable, as recognized by organizations such as the UN. This again can be attributed to online advertising, which has been the main financier to the open web. However, recent trends of paywalling information and services on the web are creating imminent dangers to such open model of the web, inhibiting access for the economically vulnerable, and eventually creating digital segregation. In this paper, we argue that this emerging model lacks sustainability, exacerbates digital divide, and might lead to collapse of online advertising. We revisit the ad-supported open web business model and demonstrate how global users actually pay for the ads they see. Using data on GNI (gross national income) per capita and average paywall access costs, we established a simple income-paywall expenditure gap baseline. With this baseline we show that 135 countries with a total population estimate of 6.56 billion people cannot afford a scenario of a fully paywalled web. We further discuss how a mixed model of the so-called “premium services” creates digital segregation and poses danger to online advertising ecosystem. Finally, we call for further research and policy initiatives to keep the web open and more inclusive with a sustainable business model.\end{abstract}
\begin{document}
\maketitle
\section{Introduction}
The advent of the World Wide Web has revolutionized the way we access and share information, leading to an enormous amount of accessible knowledge and opportunities for individuals and societies globally. The open model of the web which allows anyone with an internet connection to access and contribute information for free, has been instrumental in creating a global information ecosystem that has benefited billions of users worldwide \cite{chaqfeh2023towards-1}. This open model of the web has provided unprecedented opportunities for the most vulnerable members of society, enabling them to access educational resources, connect with communities, and participate in the digital economy and social development \cite{Vodafone-7}. It is important to note that getting access and staying connected to the internet by itself is an expensive challenge for most parts of the world \cite{papadopoulos2018cost-21}. Recognizing this challenge (also referred to as first level digital divide) various efforts have been made by governments, industry, and research bodies to bring internet connection widely accessible\cite{singh2017leveraging-43}. The UN has identified ICT access as one of sustainable development goals \cite{Vodafone-7}. Researchers have also studied approaches to address the second challenge, such as creating low-cost ICT devices, creating innovative solutions to make internet reachable to farthest places, and lowering the user cost of web traffic and improving user experience with low-speed internet\cite{habib2023framework-40}. Despite the many challenges such as high cost of connecting to internet \cite{papadopoulos2018cost-21}, online privacy \cite{papadopoulos2017if-19} and security issues \cite{sakib2015automated-26}, the ad-supported web has generally been free (arguably) and open without requiring payment beforehand, until recently. 

The primary financier of this open web has been online advertising \cite{greenwood2021you-2}. As such, online advertising has enabled content creators and service providers to monetize their offerings and sustain their businesses, providing global internet users with a wealth of free information and online services. However, as we will see in the rest of the paper, recent trends towards paywalling information and services on the web pose imminent dangers to this open model.

The multifaceted benefits of the open web is enormous and the advantages it has been offering to the least privileged has been unprecedented. Multiple reports have underlined its transformative power such as its effect on bridging employment inequality \cite{hjort2019arrival-9}, reducing the proportion of households(particularly women)  below the poverty line \cite{bahia2020welfare-10}, and helping eradicate illiteracy \cite{west2014reading-11}.

The emergence of paywalling an information content or part of it as a business model can be linked to several factors, including the high costs of producing and operating a large-scale information content and service, and the ad-tech shift towards highly targeted advertising and entailed user technologies to circumvent it \cite{nithyanand2016adblocking-4}. While these emerging paywalls may seem to provide short-term financial gains for content platforms and service providers, they exacerbate the digital divide and lack sustainability as a business-model in the long term.

In this paper, we raise concerns about the sustainability and inclusivity of the emerging trend towards paywalled content and services on the web. We highlight dangers of this emerging challenge contending that a fully paywalled web is unsustainable and poses significant risks to entire advertising-supported ecosystem. After covering some of important related work in the area, we begin with revisiting the ad-supported open web business model and demonstrate how global users actually pay for the ads they see. Using results from previous research \cite{papadopoulos2020keeping-8} and the GNI per capita data, we then discuss how cost of accessing a small portion of content on internet is impractical for most part of the world. Next, we discuss how a mixed model of premium services creates digital segregation and poses a danger to the online advertising ecosystem. With these points we aim to shed light on how this emerging business model exacerbates existing inequalities and creates new ones. Then, we proceed to call for research and policy efforts towards creating ethical and desirable advertising systems that can keep the web open and more inclusive. Finally, we conclude the paper pointing limitations of our work and suggesting potential directions for future work.

\section{Related Work}
While online advertising has been widely acknowledged as backbone of the open web \cite{Glenday_2020a-22,kassa2018large-23} enabling free global information dissemination, its modus operandi has always been the subject of user dissatisfaction \cite{chua2020threats-27}. The dissatisfaction primarily emerges from embedded adverts between user interactions that degrade user experience \cite{miroglio2018effect-15, zeng2021makes-16}, and the increasing incorporation of sophisticated tracking mechanisms. Advertisers use tracking techniques to collect user activity data with the aim of bringing most relevant ads to the users \cite{olejnik2013selling-28,budak2016understanding-17}. In relation to this, multiple open platforms serving billions of users have been a subject of several scrutiny \cite{matz2020privacy-13,patnaik2023tiktok-18}.
A particular example is the Cambridge Analytica case \cite{gonzalez2019global-12}, that attempted to influence US presidential election analysing user behavior, which was widely covered by mainstream media resulting in global awareness on privacy. Even though the Cambridge Analytica study was performed on Facebook, it is worth to note that the method can theoretically be replicated to other social platforms such as X (formerly Twitter) and TikTok. Privacy risks associated with online advertising have been widely documented \cite{gonzalez2021unique-29,leon2015privacy-30}. The advertising monetary value of user data has also been documented in multiple studies \cite{papadopoulos2017if-19,marciel2016value-20}. While these studies showed the risks and undesirable sides of online advertising and collecting user data, other results have shown how proper handling of user data such as via anonymization and aggregation can effectively be used for societal benefit \cite{garcia2018analyzing-31,giurgola2021mapping-32,rama2020facebook-41}. Furthermore, multiple works have proposed to improve user interaction with online advertising with approaches such as bringing transparency and improving user privacy \cite{parra2017myadchoices-24,fiesler2018we-25}, and ad profit sharing with users \cite{chen2008sharing-33}. A few research works have explored impact of advertising on ad-supported open web \cite{shiller2017will-34,budak2016understanding-17,papadopoulos2020keeping-8}. Varian \cite{varian2022advertising-3} studied interplay between advertising costs and product prices and has shown that when advertisers cut back on ad spending, they do not necessarily change product prices accordingly. A paper by Shiller et al. \cite{shiller2017will-34} studied the compounded impact of ad-blocking on revenue concluding that ad blocking poses a substantial threat to the ad-supported web. A more recent article of high relevance to our discussion is a paper by Papadopoulos et al. \cite{papadopoulos2020keeping-8} which studies prevalence of paywalls and discussed their impacts. Their findings show that paywall use has been increasing, and the median cost of an annual paywall access is 108 USD per site.

\begin{figure*}[ht]
\centering
\includegraphics[width=0.7\textwidth]{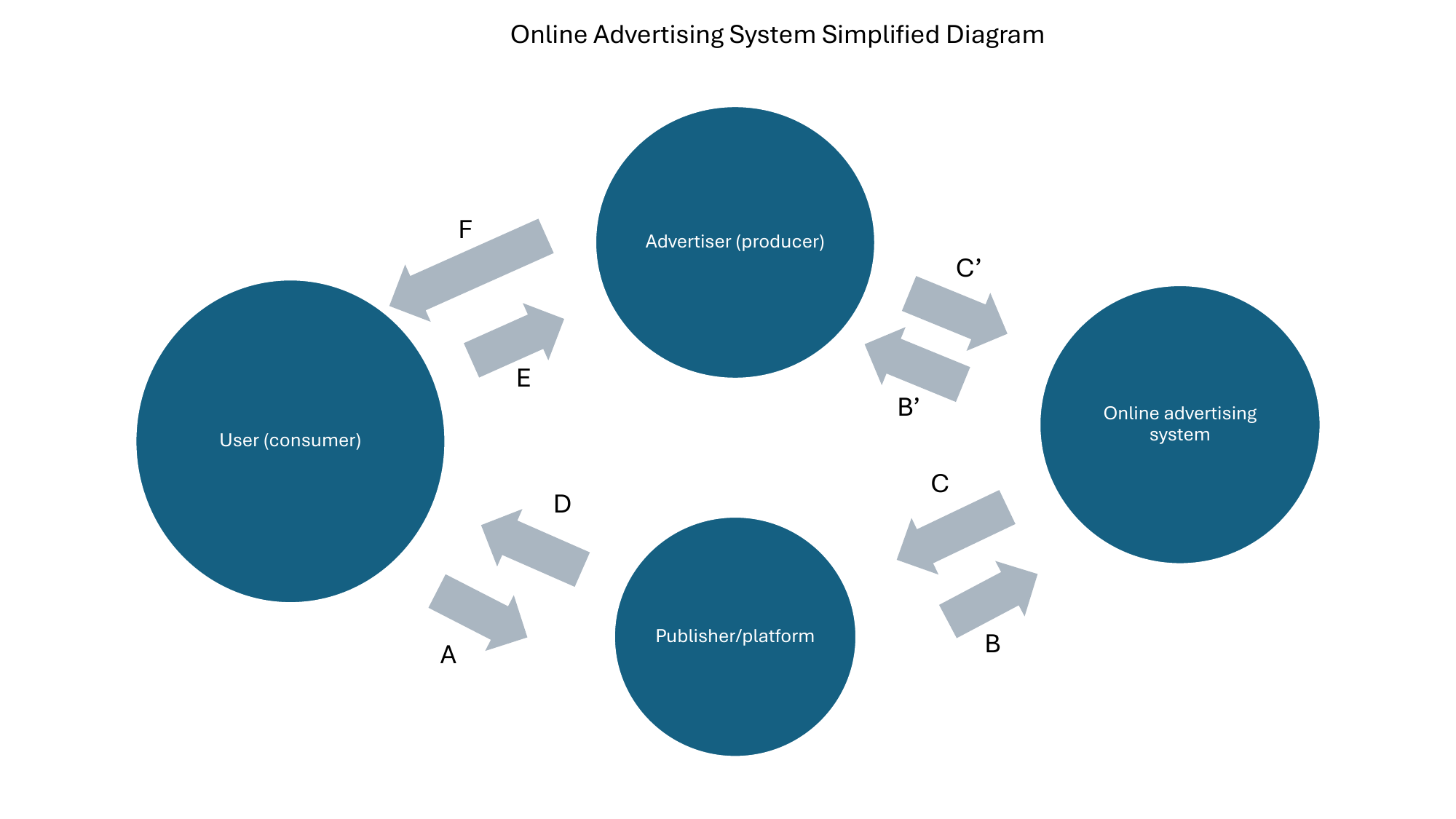}
\caption{Diagram Revisiting ad-supported Web: First,
advertiser assigns a total cost (TC) to a product/service by adding the advertising dollar (Ca) to the initial cost of a product/service. When a user visits ad-supported publisher/platform (arrow A), the platform contacts ad-system (arrow B), which in turn performs RTB (arrow B') that leads to winning advertiser sending back ads and a dollar amount (arrow C'). the advertising system then sends the winning ad along with a fraction of the dollar amount to the ad-slot on the platform (arrow C). The ad is presented to the user (arrow D). Up on conversion, user eventually pays the dollar amount for ad they saw (arrow E) plus Cps, finally product/service is delivered to user (arrow F).}
\Description{Diagram Revisiting ad-supported Web}
\end{figure*}

\section{Revisiting ad-supported Internet}

A well-known architectural presentation to discuss modern day online advertising systems involves putting advertisers and end users on the two extremes that are connected via online advertising system as intermediary. In this popular scenario the online advertising system composed of parties like supply side platform, ad exchange, and demand side platforms tracks user activities to gather data that can be used during  Realtime Bidding (RTB). When a user visits ad-supported platform RTB takes place, where potential advertisers bid for ad-slot to be displayed as part of the content presented to the end user \cite{zhang2014optimal-37}. To build the most optimal matching between advertisers and ad-slots to each individual user advertising system uses a variety of sophisticated techniques that heavily make use of personal data \cite{speicher2018potential-36}. Though it is widely known that user personal data is a tradable commodity, the fact that end users actually pay for the ads they see is often overlooked.

Online advertising is a multibillion-dollar business. This multibillion-dollar budget comes from advertisers which are producers of goods and services that aim to reach their potential clients or customers. Companies spend a significant portion of their revenue to advertising, the average ad spending budget of small businesses ranges between 4-6 percent of their gross revenue, while ad spending by some companies rises up to 20\% \cite{Frenkiel_2023-38}. 

Let's take a scenario of company X getting its customers exclusively via online advertising. If gross revenue of company X (i.e. total sum of products and services sold) is R, then a portion of this revenue goes to online advertising Ca. Let X's target cost after advertising be Cps. 

\noindent Then,
\begin{equation}
    Cps = R-Ca
\end{equation}
                 
This gross revenue, R, is the sum of total cost (TC) paid by X's clients/customers.

\begin{equation}
    TC =Cps + Ca
\end{equation}
\begin{equation}
TC = \sum_{i \epsilon items} \sum_{j \epsilon users} \left ( Cps_{(j)}^{(i)} + Ca_{(j)}^{(i)} \right )
\end{equation}
With this second equation, we can say that the advertising budget is practically paid by end users. We can diagrammatically see this model on Figure 1.  
The diagram can be summarized as follows: initally advertiser assigns a total cost (TC) to a product/service by adding the advertising cost (Ca) to the initial target cost of a product/service. The advertiser places this ad with advertising dollar Ca on RTB. The advertiser then sends the winning ad through the publisher/platform paying a fraction of Ca to the platform. Up on conversion, user eventually pays for ad they saw. This visualization demonstrates that the advertising cycle can sustainably continue as long as the system is healthy and end users find products/services advertised to them relevant and appealing. It also suggests us to postulate that end users practically pay for the ads they are exposed to. 

\begin{figure*}[ht]
\centering
\includegraphics[width=0.85\textwidth]{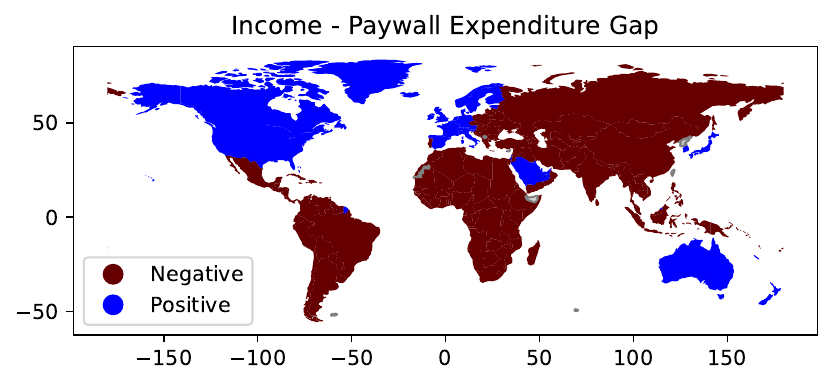}
\caption{IPEG, income paywall expenditure gap (blue indicates IPEG value $>$ 0, dark red indicates IPEG$<$0) demonstrating majority of countries having negative IPEG value indicating that paywall budget cost is already higher than the average before tax income of a country's citizens.}
\Description{IPEG, income paywall expenditure gap}
\end{figure*}
\section{Pay-Walling the Internet Exacerbates Digital Segregation}
Given there are over a billion websites in the world today, users are faced with multiple alternatives of varying degrees of utility for their digital needs. This information overload also exposes users to multiple challenges such as disinformation \cite{kumar2016disinformation-44}, and malware \cite{sakib2015automated-26}.
Services such as Google AdWords and SimilarWeb provide categorization of websites into interest groups such as books, music, business services, adult, games, etc. SimilarWeb\footnote{https://support.similarweb.com/hc/en-us/articles/213856729-Top-Websites} for instance provides 250 such categories. It is challenging to analyze and categorize each individual website to decide which platform can be used as a replacement (representative) for other websites in its category. For example, CNN and Fox News fall under news and media category, but it is known that one cannot replace the other \cite{flamino2023political-39}. An interesting example of website uniqueness is the case when the Seattle times temporarily had to lift its paywalls during the 2014 Oso mudslide recognizing that it was the main platform covering the disaster on its website \cite{ananny2016drop-6}.

However, for our discussion we will assume that a user can satisfy their information needs with one website from each category (i.e. user can live with only 250 websites). According to results from Papadopoulos et al. \cite{papadopoulos2020keeping-8}, median annual paywall access is 108 per site, this gives that in a fully paywalled scenario a user will need to budget 27,000 USD (i.e. 108 USD*250) just to get past paywalled services. Taking the GNI (Gross national income) per capita, GNIpc, of countries from world bank, we can estimate a baseline income-paywall expenditure gap, IPEG, for users per country.

 \begin{equation}
    IPEG = GNIpc-PWAC
\end{equation}

For paywall business model to be sustainable, the IPEG has to be positive ( bare minimum). On Figure 2, we show that IPEG is negative on 135 countries (amounting to 6.56 billion people). Note that some of those countries don’t even have online payment system even if a few of users can afford to pay for the paywalls. Also, it is important to remember that even in those countries with blue IPEG, a significant portion of the population will have difficulty allocating paywall budget.

Now let’s discuss the hybrid model which is spearheaded by large platforms such as YouTube, Twitter (X), and TikTok \footnote{https://techcrunch.com/2023/02/27/social-media-apps-adopting-subscription-models/}. Data on of some of the most popular platforms and their premium fees is presented on appendix A. One of the main incentives to go premium on such platforms is ad-free experience or a service with fewer ads. Hybrid model gives rich people option to avoid watching ads, get higher quality information and experience. However, as we saw previously, targeted advertising via RTB works by finding potential customers for each advertiser.  If certain groups of advertisers can’t find their target users they will avoid investing in online advertising (e.g. advertisers of cars, luxury watches, villas, high end phones will pull their advertising budget for lack of reaching potential customer) and hence reducing the market capitalization of total advertising ecosystem. In other words, if we rearrange equation 3 based on income,  rich and affluent groups will disappear from equation 5.
\tiny
\begin{equation}
    TC =(Cps + Ca)^{rich} + (Cps + Ca)^{affluent} + (Cps + Ca)^{poor}
\end{equation}
\normalsize
Note that advertisers pulling from platforms has immediate effect with potential to kill the company (advertisers vs X (formerly Twitter), for example\footnote{ https://time.com/6340905/elon-musk-advertiser-boycott-x/}). Reduced advertising economy will discourage content generators and online service providers from serving the poor, specifically people in least developed regions will be at a high risk. On the other hand, economically vulnerable people in developed regions will be unable to access paywalled content and hence will be digitally segregated getting less quality digital content, making them less informed, and less competent in general.

\section{Conclusion}
\subsection{We Need More Desirable and Ethical ads}
So far we have discussed the important role ads play in keeping the web open, and the impracticality of the emerging premium-paywall business models. We have also raised some of the concerns that make advertising undesirable part of the web. However, there is not alternative business model that is viable and as scalable as online advertising that can drive an open and inclusive global scale information exchange. If we address its undesirable issues,  Online advertising can be the most viable information super-highway that connects producers with consumers, or act as a mechanism to deliver timely information to the masses as it was proven during COVID-19 pandemic \cite{breza2021effects-45}. Our position is that with enough research and strong policy, there is a possibility to device an advertising framework that balances the needs of different stakeholders while ensuring the sustainability and inclusivity of the web. Rather than shifting towards paywalled services, our focus should be on fixing and refining this incumbent infrastructure to make advertising a desirable feature of the web.

\section{Limitations and Future Work }
While we cover several aspects of this emerging issue in this short paper, our approach has several limitations. First, in our IPEG calculation we didn't take into account other expenses which made our baseline highly generous. Second, we relied on SemanticWeb categorization, a more detailed and sophisticated content and traffic analysis could result in a more accurate number of representative websites. Finally, we have not included the possible solutions to improve online advertising. Future work could capitalize on these limitations.

\section{Acknowledgments}
Author(s) would thank reviewers for their valuable review and insightful comments.

\bibliographystyle{ACM-Reference-Format}
\bibliography{bibfile}

\appendix
\clearpage
\section{Appendix A: premium costs of some of the popular platforms}

\begin{table}[h]
\centering
\resizebox{\textwidth}{!}{%
\begin{tabular}{|l|l|l|l|}
\hline
\textit{\textbf{Platform}} & \textit{\textbf{Monthly}} & \textit{\textbf{Annual}} & \textit{\textbf{Name}} \\ \hline
youtube.com   & 13.99 & 167.88 & YouTube premium \\
instagram.com & 11.99 & 143.88 & verified  \\
twitter.com & 16 & 192.0 & premium+ \\
x videos.com & 9.99 & 119.88 & premium \\
tikTok.com & 4.99 & 59.88 & premium \\
reddit.com & 5.99 & 71.88 & premium \\
openai.com & 20 & 240 & plus \\
netflix.com & 22.99 & 275.88 & premium \\
office.com & 6.99 & 83.88 & personal \\
Zoom.us & 13.33 & 159.96 & pro\\ \hline
\textit{Total} & \textit{126.26} & \textit{1515.12} & \\ \hline
\end{tabular}%
}
\caption{Paywall costs of some of the popular websites from top 50 websites based on similarweb.com. Note that some of the platforms offer discounts when purchased in bulk or if paid for the full year plan.}

\end{table}
\end{document}